\documentclass[12pt]{article}
\usepackage{graphicx}

\def\cmu{Department of Physics\\
Carnegie Mellon University, 5000 Forbes Ave., Pittsburgh, PA 15213}
\def\support{\footnote{
  Work supported by the DOE under Contract No. DESC0010118.}}

\def\Title#1{\begin{center} {\Large #1 } \end{center}}
\def\Author#1{\begin{center}{ \sc #1} \end{center}}
\def\Address#1{\begin{center}{ \it #1} \end{center}}

\newenvironment{Abstract}{\begin{quotation}  }{\end{quotation}}
\newenvironment{Presented}{\begin{quotation} \begin{center} 
             PRESENTED AT\end{center}\bigskip 
      \begin{center}\begin{large}}{\end{large}\end{center} \end{quotation}}
\def\Acknowledgements{\bigskip  \bigskip \begin{center} \begin{large}
             \bf ACKNOWLEDGEMENTS \end{large}\end{center}}

\def\beq{\begin{equation}}
\def\eeq#1{\label{#1}\end{equation}}
\def\eeqn{\end{equation}}

\def\beqa{\begin{eqnarray}}
\def\eeqa#1{\label{#1}\end{eqnarray}}
\def\eeqan{\end{eqnarray}}

\let\bar=\overbar

\def\Dslash{\not{\hbox{\kern-4pt $D$}}}
\def\dslash{\not{\hbox{\kern-2pt $\del$}}}

\def\msb{{\bar{\ssstyle M \kern -1pt S}}}

\begin{document}
\begin{titlepage}

\vfill
\Title{Quantum Correlated Charm at Threshold and Inputs to 
Extractions of $\gamma$ from $B$ Decays}
\vfill
\Author{Roy A. Briere\support}
\Address{\cmu}
\vfill
\begin{Abstract}
The basic physics of quantum-correlated $D^0, \bar{D}^0$ pairs produced 
at threshold via decays of the $\psi(3770)$ is introduced.   
The connection to extractions of the CKM angle $\gamma(\phi_3)$ 
from $B$ decays is emphasized throughout.  
Recent quantum correlation results from BESIII and CLEO-c 
are then summarized before closing with a discussion of selected issues.  
\end{Abstract}
\vfill
\begin{Presented}
The 8th International Workshop\\
on the CKM Unitarity Triangle (CKM 2014) \\
Vienna, Austria, September 8-12, 2014
\end{Presented}
\vfill
\end{titlepage}
\def\thefootnote{\fnsymbol{footnote}}
\setcounter{footnote}{0}

\section{Introduction}

An $e^+e^-$ collider running at the $\psi(3770)$ resonance near 
open-charm ($D\bar{D}$) threshold will produce entangled neutral 
$D$ meson pairs.  
This entanglement, or quantum coherence, leads to a variety 
of interesting effects and in particular allows convenient 
access to certain relative phases.  
This phenomenon has been known for some time \cite{GoldhaberRosner_1977} 
and by now consequences have been explored in detail  
\cite{Xing_1997, GGR_2001, AtwoodPetrov_2005, AsnerSun_2006}.  
It is quite interesting to observe these EPR-like quantum correlation 
effects in an HEP experiment.  

One major goal of quark flavor physics involves over-constraining 
the CKM mixing matrix in an effort to find evidence for new physics.  
Measurements of the CKM angle $\gamma$ (or $\phi_3$) may {\it in principle} 
be made with negligible theoretical uncertainty using $B \to DK$ decays 
\cite{BrodZupan_2014}.  
Indeed, several experiments are already exploiting this via analyses 
of a wide variety of related decay chains \cite{HFAG_gamma}.  

In order to perform clean extractions of $\gamma$, one must 
avoid introducing unnecessary model-dependence in the analyses.  
Data from charm threshold can be used to to avoid such pitfalls, 
as we will see below, by directly measuring the strong-phase quantities 
that the $B$ decay analyses need.  
Such external inputs from charm are useful in avoiding ambiguities, 
simplifying analyses and reducing uncertainties.  
They also have the virtue of replacing hard-to-evaluate uncertainties 
from model dependence with clearer and largely statistical uncertainties 
from the threshold charm results.  

Quantum-correlated effects appear in three places: (i) correlated 
charm at threshold, (ii) $B \to DK$ analyses with common $D$ final states, 
and (iii) charm mixing.  Most of this review concerns using (i) as an input 
to (ii).   At the end, we will briefly cite discussions of using 
(i) as an input to (iii), and (iii) and an input to (ii).

\section{Charm at Threshold and Quantum Coherence}

Production of the $\psi(3770)$ state is followed about half of the time 
by decay into a neutral $D$ meson pair which is entangled as: 
\begin{equation}
 \psi(3770) \;\to\; \frac{1}{\sqrt{2}}\;
    \left[\;  D^0(+z) \bar{D}^0(-z) \;-\; \bar{D}^0(+z) D^0(-z) \;\right] 
\label{Eqn:one}
\end{equation}
or, using the $CP$ eigenstate combinations 
$D_{CP\pm} = [D^0 \pm \bar{D}^0]/\sqrt{2}$
\begin{equation}
   \psi(3770) \;\to\; \frac{1}{\sqrt{2}}\;
    \left[\;  D_{CP-}(+z) D_{CP+}(-z) \;-\; D_{CP+}(+z) D_{CP-}(-z) \;\right] 
\label{Eqn:two}
\end{equation}
In both cases, the (arbitrary) center-of-mass decay axis is labelled 
as $\pm z$.  
Experiments measure various combinations of rates, such as the total rate 
for one $D$ decay final state (inclusive ``single tags'') 
or for specific pairs of $D$ decay modes (exclusive ``double tags''). 
These rates display some very interesting effects arising from 
quantum coherence and the resulting interference.  
In particular, the double-tag rates have useful sensitivities 
while single-tag rates provide useful for normalizations.  

One convenient way to introduce interference effects involves observing 
one $D$ meson decaying to a $CP$ eigenstate; this projects the other meson 
into a coherent $D^0, \bar{D}^0$ superposition with the opposite $CP$, 
as displayed in Eqn. \ref{Eqn:two}.  
If that second $D$ decays to a mode accessible to both the $D^0$ and 
$\bar{D}^0$ components of the state, then there will be interference.  
One can even change the sign of the interference term by choosing 
$CP-$ instead of $CP+$ eigenstates for the first decay.  
While this is a nice example, it is important to realize that 
the participation of $CP$ eigenstates is not essential.  
There is coherence in the $D$-pair wavefunction no matter which basis 
is chosen and this leads in general to interference effects.  

It is useful to classify some key types of $D$ meson decays.  
Cabibbo-favored (CF) decays result from $c \to sW^+, W^+ \to u\bar{d}$ 
transitions, producing one\footnote{Or, rarely, producing $\bar{K}K\bar{K}$, 
but such decays are not relevant for us here.} $\bar{K}$ meson 
(a $K^-$ or $\bar{K}^0$).
Singly-Cabibbo-suppressed (SCS) decays result from 
$c \to sW^+, W^+ \to u\bar{s}$ and $c \to dW^+, W^+ \to u\bar{d}$ 
transitions, resulting in an even number (0 or 2) of kaons.  
Doubly-Cabibbo-suppressed decays (DCSD) are from the process 
$c \to dW^+, W^+ \to u\bar{s}$ resulting in a ``wrong-sign'' $K$ meson.   
Actually, ``wrong-strangeness'' is a more appropriate terminology; 
however when the single kaon produced is electrically charged, 
then CF (DCSD) decays of $D$ mesons produce a $K^-$ ($K^+$), and the 
opposite for $\bar{D}$ decays.  There is a suppression of other decays 
relative to CF controlled by the Cabibbo angle, $\theta_C$: SCS decays 
are suppressed {\it in amplitude} by one power of $\tan\theta_C \sim 0.22$, 
while DCSD suffer from two powers of this same factor.  

In addition to the Cabibbo factors corresponding to a given $D$ decay, 
certain other properties are also worth noting.  
Hadronic decays refer to the cases where all decay products are mesons, 
while semileptonic means that there is a charged lepton and neutrino 
($e \nu_e, \mu \nu_\mu$) produced along with one or more hadrons.  
Some final states are symmetric (self-conjugate) with respect to 
particle-antiparticle exchange, while others are not.  
Also, some final states are ``flavored'', meaning that one can infer 
whether the initial state contained a $c$ or $\bar{c}$ quark.  
Such flavor-tagging is valid only at the time of decay, due to 
$D^0-\bar{D}^0$ mixing.  
Furthermore, it is only exact for semileptonic decays: 
for CF hadronic decays, flavor tagging is contaminated by 
the rarer DCSD processes.  

The various $D$ decays relevant to our discussions are summarized in 
Table \ref{Tab:modes}.  
We note that it is possible to be neither self-conjugate nor flavored 
but it is not possible to be both.  

\begin{table}[htb]
\label{Tab:modes}
\begin{center}
\begin{tabular}{|l|l|l|}
\hline
{\bf Type} & {\bf Examples} & {\bf Amplitude(s)} \\
\hline \hline
{\it Flavored} & & \\
\hline
$-$ Flavored semileptonic & $K^-e^+\nu, K^-\mu^+\nu$ & Pure CF \\
$-$ Flavored hadronic  & $ K^- \pi^+, K^-\pi^+\pi^0, K^-\pi^+\pi^+\pi^-$ 
                                                  & CF + DCSD \\ 
\hline \hline
{\it Self-conjugate} & & \\
\hline
$-$ 2-body $CP$ eigenstate & $K^-K^+, \pi^+\pi^- , K_S\pi^0$ & SCS \\
$-$ Multi-body I  & $K_S h^+h^-, K_Lh^+h-$ & CF + DCSD \\
$-$ Multi-body II & $ K^+K^-\pi^+\pi^-, \pi^+\pi^-\pi^0$ & SCS \\
\hline \hline
{\it Neither} & $K_S K^- \pi^+,K_S K^- \pi^+\pi^0 $ & SCS \\
\hline \hline
{\it Both} & {\it - not logically possible -} & \\
\hline
\end{tabular}
\caption{A summary of relevant classes of $D$ decays; 
$h$ indicates  $\pi$ or $K$.  
``Neither'' and ``both'' refer to being or not being 
flavored and self-conjugate.}
\end{center}
\end{table}

\section{Multi-Body Coherence Factors}

Consider the interference of two two-body amplitudes: 
\begin{equation}
 |{\cal A}_1 + {\cal A}_2|^2  \,=\,  
  |{\cal A}_1|^2 + |{\cal A}_2|^2 + 2 |{\cal A}_1| |{\cal A}_2| \cos\delta
  \,=\, A_1^2 + A_2^2 + 2 A_1 A_2 \cos\delta
\end{equation}
where 1 and 2 label, for example, the CF and DCSD $D^0 \to K^\mp\pi^\pm$ 
amplitudes, and $\delta$ is their {\it relative} phase.  The second equality 
uses the magnitudes of the amplitudes, $A_i = |{\cal A}_i|$.  
The generalization of this to multi-body decays results in the 
introduction of Atwood-Soni coherence factors \cite{AtwoodSoni_2003}.  
In this case, one has to integrate over the Dalitz plot.  
After doing so, these coherence factors allow one to construct 
an expression very similar to the two-body relation.  

We will need to distinguish between the {\it complex} amplitudes 
at one point in phase space (denoted by $x$)$, {\cal A}_i(x)$, 
and their {\it real} phase-space-averaged integrals, $A_i$, 
defined via $A_i^2 = \int \,dx\, |{\cal A}_i|^2 $.  
These generalize the simple $A_i = |{\cal A}_i|$ relation 
of the two-body case.  
For multi-body cases, we have: 
\begin{equation}
\label{Eqn:simple}
 \int dx \;\; |{\cal A}_1 + {\cal A}_2|^2  \,=\,  
     A_1^2 + A_2^2 + 2 R A_1 A_2 \cos\delta 
\end{equation}
Here, the coherence factor is given by $R$ and $\delta$ which summarize 
the net effect of amplitude and phase variation across the Dalitz plot.  
Using $D^0 \to K^\mp(n \pi)^\pm$ decays as an example: 
\begin{equation}
  R_{Kn\pi} \, e^{-i\delta_{Kn\pi}} \,=\, 
     \frac{\int \,dx\, {\cal A}_{K^-(n\pi)^+} \, {\cal A}_{K^+(n\pi)^-}}
           {A_{K^-(n\pi)^+} \, A_{K^+(n\pi)^-}}
\end{equation}
where we have simply taken the ratio of the correct cross-term to 
the simplified average form used in Eqn. \ref{Eqn:simple}.  
In a two-body decay $R=1$ and there is only a relative phase $\delta$ 
which controls the Dalitz-averaged CF-DCSD interference.  In the multi-body 
case, there are {\it two} real parameters remaining: $R, \delta$.  
But we can also use these parameters to discuss decays beyond 
$K^-(n\pi)^+$.  

In general, we expect the parameter $R<1$ due to two effects.  
First, the $x$-dependent phase relative phase 
$\arg({\cal A}_2^*(x){\cal A}_1(x))$ varies as we integrate.  
Second, the amplitudes don't track each other in magnitude 
across phase space; i.e., the local amplitude ratio 
$|{\cal A}_2(x)/{\cal A}_1(x)| \ne 1$ .

Viewed as polar coordinates, $R, \delta$ lie on or inside the unit circle.  
It is also possible to consider a Cartesian basis, 
$(c, s) = (R \cos\delta, R\sin\delta)$.  
One can also obtain results in separate sub-regions, or bins, 
of the Dalitz plot, labeled by a subscript $i$.  
As we will see below, one typically uses (for various reasons) 
$R, \delta$ for $K^-(n\pi)^+$ and several bins of $c_i, s_i$ for 
$K_S\pi^+\pi^-$ and related modes.

\section{Quantum Correlations for Pedestrians}

The simplest quantum correlation effect in $\psi(3770)$ decays 
involves both mesons decaying to $CP$-eigenstates.  Like-$CP$ 
($++, --$) combinations are forbidden, while opposite-$CP$ ($+-, -+$) 
are enhanced by two, as one can see directly from the wave function 
in Eqn. \ref{Eqn:two}.  
However, as noted above, interference effects are quite general.  
We denote the decay amplitudes as 
${\cal A}(     D^0   \to F) =      {\cal A}_F$ and 
${\cal A}(\bar{D}^0  \to F) = \bar{{\cal A}}_F$.  
For the real, averaged amplitudes defined above, we have 
$\bar{A}_F = A_{\bar{F}}$ and $\bar{A}_{\bar{F}} = A_F$.  
Then, the decay width for a double-tag final state 
to modes $F,G$ is given by: 
\begin{equation}
  \Gamma_{FG} \,=\, \Gamma_0 
 \left[ 
    A_F^2 \bar{A}_G^2 + \bar{A}_F^2 A_G^2 
       - 2 A_F \bar{A}_F A_G \bar{A}_G R_F R_G \cos(\delta_G - \delta_F)
 \right]
\end{equation}
If, for example, we take $F = K^-\pi^+\pi^0$, $G = K^-\pi^+\pi^+\pi^-$, 
then factoring out the larger amplitudes gives
\begin{equation}
 \Gamma_{FG} \,=\, \Gamma_0 A_F^2 A_G^2 
 \left[ 
    r_F^2 + r_G^2 - 2 r_F r_G R_F R_G \cos(\delta_G - \delta_F)
 \right]
\end{equation}
Here, $r_{F,G} = \bar{A}_{F,G}/A_{F,G}$, defined such that $r \le 1$.  
For other cases, like opposite-sign $K^-(n\pi)^+$ vs. $K^+(n\pi)^-$, 
one would replace $r_F^2 + r_G^2$ by $1 + r_F^2 r_G^2$ to keep $r \le 1$.    
This form is considerably simplified in many other cases.  For example, 
$r = \pm 1$ for $CP$ eigenstates, $r=0$ for semileptonic decays.  
But $r \simeq \tan^2 \theta_C \simeq 0.05$ for CF+DCSD cases.  
For $K^\mp(n\pi)^\pm$, $R, \delta$ are the Atwood-Soni 
coherence factors which are a priori unknown, except that 
$R=1$ for $n=1$ ($K^\mp\pi^\pm$).  \
They are trivial for $CP$-eigenstates ($R=1, \delta = 0, \pi$)  
and for semileptonic decays ($R=0$).   

Note that interference in general is only sensitive to 
$Re(e^{-i\delta}) = \cos\delta$.  
But $\delta$ here is a {\it difference} between the two relative phases 
for the two final states of the double-tag.  
If one decay mode has a trivial phase, then is it true that 
one will only be sensitive to $\cos\delta$ for the other non-trivial 
amplitude.  But if both decays have non-trivial phases, one gets 
\begin{equation}
 \cos(\delta_G - \delta_F) \,=\, 
   \cos\delta_G \, \cos\delta_F + \sin\delta_G \, \sin\delta_F
\end{equation}
One can measure enough observables to separately determine 
both the $\sin$ and $\cos$ terms and thereby recover sensitivity 
to $\sin\delta$ (and hence the sign of $\delta$).  
This can be achieved in the obvious way by employing two different 
decay modes, for example different $n$ with $K(n\pi)$ modes.  
But it can also be done with different portions of phase-space 
in a multi-body decay which act as independent modes.  
Note that when using $c_i, s_i$, we must recall that these arise as 
Cartesian coordinates of some binned $Re^{-i\delta}$, 
and not $e^{-i\delta}$.  Hence, $s_i^2 \ne 1-c_i^2$: 
there are still two independent degrees of freedom.  


\section{\boldmath Charm Threshold and $B$ Physics}

There are a number of methods to extract the CKM angle $\gamma$ 
from $B \to DK$ decays, an ``alphabet soup'' of acronyms, distinguished 
largely by the specific $D$ decay involved.  The key is to exploit 
a final state which is common to both $D^0$ and $\bar{D}^0$ decays.  
We first give a brief summary of methods, and then discuss what charm 
data at threshold can do to help such analyses.  

The pioneering GLW method \cite{GL_1991, GW_1991} uses $D$ decays 
to a $CP$ eigenstate.   
The ADS method \cite{ADS_1997, ADS_2001} employs modes that are CF and 
DCSD; this helps to balance the overall amplitudes to maximize interference 
effects.  
The GGSZ method \cite{Bondar_2002, GGSZ_2003, BP_2006, BP_2008} 
extends analyses to Cabibbo-favored self-conjugate multi-body modes.  
We also note the infrequently mentioned GLS paper \cite{GLS_2003}, 
which proposed using SCS decays to non-$CP$-eigenstates (such as 
$K^*K$, which decays to $K_SK\pi$).  
The utility of $B \to D^* K$ in addition to $B \to DK$ 
has also been explored \cite{BondarGershon_2004}.  
As one sees, $B$ factories (BaBar, Belle, LHCb, and soon BelleII) 
have many choices of $D$ decay modes to use when studying $\gamma$ 
\cite{CKM14_gamma}.  

We now quickly survey the charm physics within each $\gamma$ method.  
Semileptonic $D$ decays are impractical for $B$ physics, 
but they are useful for charm threshold work.  
Since there is no interference, they are useful for normalization.  
For the $CP$-eigenstates of GLW, the strong phase differences between 
the $D^0$ and $\bar{D}^0$ amplitudes are always trivial: $0$ or $\pi$.  
There is no need for input from charm threshold studies, but these 
final states are very useful as one half of double-tag combinations 
in our threshold charm analyses.  

For the ADS method, there is a non-trivial strong phase between the 
CF and DCSD amplitudes.  
Threshold data can provide the necessary Dalitz-averaged 
Atwood-Soni coherence factors (two parameters in general, 
or in the two-body $K\pi$ case just one relative phase).  
These modes are also sometimes used as normalization since they 
are easier to reconstruct than semileptonic decays, but one must 
be careful to correctly account for the DCSD effects.  

Multi-body self-conjugate modes are the basis of the GGSZ method, 
and threshold can provide strong-phase information similar 
to the previous case.  

In cases where a multi-body state is dominated by sub-components 
that are dominantly of one $CP$ value, it is interesting to 
measure the $CP$-purity of the state: that is the fraction of 
the dominant $CP$, denoted $F_+$ for the $CP+$ fraction, etc.  

If we  examine the role of inputs measured at charm threshold 
in more detail, there are two general motivations.  
These are avoiding model dependence and accessing strong phases.  

Large and cleanly-separated samples of $D^0$ and $\bar{D}^0$ decays 
are available from $B$ factories via $D^{*+} \to D^0 \pi^+$ tagging.  
Much of the strong phase variation across the Dalitz plot may be studied 
via commonly-used fits to isobar models or related extensions.  
However, using such models introduces systematic uncertainties that 
can be avoided by measuring the quantities of interest directly 
at charm threshold.  

In addition, in some cases, there is a strong phase that is simply 
not accessible via flavor-tagged $D^0, \bar{D}^0$ samples alone.  
Imagine that we fit both CF $D^0 \to K^-\pi^+\pi^0$ 
and DCSD $D^0 \to K^+\pi^-\pi^0$ with $N$ isobar amplitudes.  
Each fit is sensitive to $N-1$ relative phases, so $2N-2$ relative phases 
are measured.  But at threshold, we can interfere these two process, 
and since all $2N$ amplitudes are involved, one is sensitive 
to $2N-1$ phases: one more.  
Said another way, in separate isobar fits, one amplitude in each fit 
is chosen as real.   But there is no linkage between the two fits 
and thus it is impossible to measure the relative phase of these 
reference amplitudes.  

For the ADS and GLS modes, both issues are relevant.  
For GGSZ modes, only the model-dependence is relevant: the relative $D^0, 
\bar{D}^0$ phase is again trivial as it is for two-body $CP$-eigenstates.  

We can summarize the experimental outputs in Table \ref{Tab:results}.  
\begin{table}[htb]
\label{Tab:results}
\begin{center}
\begin{tabular}{|l|l|c|l|}
\hline
Mode & Method & Observables & References \\
\hline \hline
$K^-K^+, \pi^+\pi^-$      &  GLW  & $-$ 
 & $-$ \\
$K^-\pi^+$                &  ADS  & $\delta$ 
 & \cite{CLEO_TQCA_2008, CLEO_TQCA_2012, BES3_Kpi_2014} \\
$K^-\pi^+\pi^0, K^-\pi^+\pi^+\pi^-$  & ADS+  & $R, \delta$ 
 & \cite{CLEO_Knpi_2009, CLEOLEG_Knpi_2014}  \\
$K_S K^- \pi^+$  & GLS & $R, \delta$ 
 & \cite{CLEO_KSKpi_2012} \\
$K_S \pi^+\pi^-, K_S K^+ K^-$  & GGSZ & $c_i, s_i$ 
 & \cite{CLEO_KSpipi_2009, CLEO_KSpipi_2010, BES3_KSpipi_2014}\\
$\pi^+\pi^-\pi^0, K^+K^-\pi^0$ & GLW, GGSZ & $F_+$ 
 & \cite{CLEOLEG_CP_2014} \\
\hline
$K^+K^-\pi^+\pi^-$ & GGSZ & $D^0, \bar{D}^0$ isobar fits  
 & \cite{CLEO_KKpipi_2012} \\
\hline
\end{tabular}
\caption{
Summary of quantities currently accessed with charm threshold data, 
arranged by mode and $\gamma$ method, with references.  }
\end{center}
\end{table}

\section{A Survey of Results from Charm Threshold}

Results are available from both CLEO-c and BESIII.  
CLEO-c analyses mainly use 0.818 fb$^{-1}$ of $\psi(3770)$ data\footnote{
The $K_SK\pi$ analysis adds in 15 fb$^{-1}$ of continuum charm data taken 
near 10 GeV, while the $KK\pi\pi$ results use 24 fb$^{-1}$ of 
such continuum charm and 0.6 fb$^{-1}$ of data taken at 4170 MeV.}, 
while BESIII uses 2.92 fb$^{-1}$.  

We begin by reviewing the CLEO-c results.  
The $K^-\pi^+$ phase was first reported in 2008 \cite{CLEO_TQCA_2008} 
and then updated in 2012 \cite{CLEO_TQCA_2012}; both versions make 
use of a complex global analysis and fit.  Using external mixing constraints, 
they find $\cos\delta_{K\pi} = 1.115^{+0.19}_{-0.17}\,^{+0.00}_{-0.08}$.  
In 2009, Atwood-Soni coherence factors were published 
for $K^-\pi^+\pi^0$ and $K^-\pi^+\pi^+\pi^-$ \cite{CLEO_Knpi_2009}.  
Two-dimensional likelihood contours are provided in the paper; 
here we simply quote $R_{K\pi\pi^0} = 0.84 \pm 0.07, 
\delta_{K\pi\pi^0} = (227^{+14}_{-17})^\circ$, and 
$R_{K3\pi} = 0.33^{+0.20}_{-.023}, \delta_{K3\pi} = (114^{+26}_{-23})^\circ$.  
The $K\pi\pi^0$ state is highly coherent, while the $K3\pi$ state is 
noticeably less coherent.  
In 2012, results for $K_SK^+\pi^-$ followed \cite{CLEO_KSKpi_2012}: 
$R_{K_SK\pi} = 0.73 \pm 0.08, \delta_{K_SK\pi} = (8.3 \pm 15.2)^\circ$; 
results were also presented for a restricted $K^*$ region.  
The quantities $c_i, s_i$ for $K_S \pi^+\pi^-$ and related modes, 
needed for the GGSZ method, have also been presented in 2009 and 2010 
\cite{CLEO_KSpipi_2009, CLEO_KSpipi_2010}.  
The first paper uses $K_L\pi^+\pi^-$ events to improve statistics, 
while the second analyzes both $K_{S,L}\pi^+\pi^-$ and $K_{S,L}K^+K^-$.  
For $K_S\pi\pi$, the key results are values of $c_i, s_i$ in eight 
bins across the Dalitz plot.  
All of the results in this paragraph have already been used 
in measurements of $\gamma$ with $B \to DK$ decays \cite{HFAG_gamma}.  

While not a quantum correlation analysis, CLEO-c also investigated 
the SCS mode $K^+K^-\pi^+\pi^-$.   
In prior studies, resonance structures were investigated, but data 
was not flavor-tagged: $D^0$ and $\bar{D}^0$ decays were mixed together.  
This made estimates of the power of this mode for $\gamma$ analyses 
uncertain \cite{RW_KKpipi_2007}.  Isobar model fits to the flavor-separated 
samples from CLEO-c greatly improved the ability to forecast this mode's 
prospects for use in extracting $\gamma$ \cite{CLEO_KKpipi_2012}.  

We now turn to several more recent results.  
The first is a ``CLEO-c legacy'' result; i.e., one performed by 
a small subset of the former collaboration using the legacy dataset.  
It is an update to the $K^-(n\pi)^+$ Atwood-Soni coherence factor 
analysis \cite{CLEOLEG_Knpi_2014}.  
The main improvements are inclusion of $K_S \pi^+\pi^-$ tags 
and use of updated external inputs (branching fractions, mixing parameters, 
$K\pi$ strong phase). 
The updated results are now 
$R_{K\pi\pi^0} = 0.82 \pm 0.07, 
\delta_{K\pi\pi^0} = (164^{+20}_{-14})^\circ$, and 
$R_{K3\pi} = 0.32^{+0.20}_{-0.28}, \delta_{K3\pi} = (225^{+21}_{-78})^\circ$.  

Shortly after the CKM2014 workshop, a new CLEO-c legacy result appeared, 
on the $CP$ purity  of the $\pi^+\pi^-\pi^0$ and $K^+K^-\pi^0$ final states 
\cite{CLEOLEG_CP_2014}.  
Specifically, the $CP+$ fraction, $F_+ = N_+/(N_+ + N_-)$, was measured, 
where $N_{+,-}$ are the normalized yields for the $CP+, CP-$ 
components of the signal decays.  These are determined using $CP$ tags 
and the knowledge the only unlike $CP$ combinations are allowed.  
They find $F_+(\pi^+\pi^-\pi^0) = 0.968 \pm 0.017 \pm 0.006$ and 
$F_+(K^+K^-\pi^0)     = 0.731 \pm 0.058 \pm 0.021$.  
The $F_+$ value for the $3\pi$ mode is close enough to 1 that it 
essentially acts like the 2-body GLW $CP$-eigenstates, with a modest 
dilution effect.  In particular, this dilution will be 
$D = (N_+ - N_-)(N_+ + N_-) = 2F_+-1$.  

BESIII has now entered the game with their larger dataset, recently 
publishing a $K^-\pi^+$ strong-phase result \cite{BES3_Kpi_2014}.  
The analysis is simpler than the related CLEO-c result, and concentrates 
on the effect of the relative $K\pi$ strong phase on the $CP$-tagged 
branching ratio asymmetry 
\begin{equation}
 A^{CP}_{K\pi} \,=\, 
  \frac{ {\cal B}(D_{CP-} \to K\pi) - {\cal B}(D_{CP+} \to K\pi) }
       { {\cal B}(D_{CP-} \to K\pi) + {\cal B}(D_{CP+} \to K\pi) }
\end{equation}
The directly observed asymmetry is $A^{CP}_{K\pi} = (12.7 \pm 1.3 \pm 0.7)\%$ .
With external inputs for mixing parameters, they then extract 
$\cos \delta_{K\pi} = 1.02 \pm 0.11 \pm 0.06 \pm 0,01$ where the last 
error is from those inputs.  The statistical error dominates 
and is about 60\% of the final CLEO-c result.  

Finally, there is a preliminary BESIII result on the GGSZ $c_i, s_i$ 
parameters for $K_S \pi^+\pi^-$ \cite{BES3_KSpipi_2014}.  
The statistical errors are improved relative to CLEO-c due to the factor 
of over 3.5 increase in the available integrated luminosity.  
As with CLEO-c, $K_L\pi^+\pi^-$ events are used to improve statistics.  
Final results for use in $\gamma$ extraction should be available soon.

\section{Selected Issues}


As extractions of $\gamma$ with these methods become more accurate, 
care must be taken to use accurate calculations of relevant rates.  
We discuss a selection of relevant issues next.  
 
With larger threshold datasets, one needs to carefully consider if inclusion 
of the $K_Lh^+h^-$ modes in the $K_Sh^+h^-$ analyses brings in unwanted 
model-dependence \cite{CLEO_TQCA_2008, CLEO_TQCA_2012, BES3_Kpi_2014}.  

Studies of the effects of both $D$ mixing \cite{MeccaSilva_1998, ASS_1999, 
SilvaSoffer_2000, GSZ_2005, BPV_2010, Rama_2014}
and $D$ $CP$ violation 
\cite{Wang_2013, MartoneZupan_2013, BGLR_2013, BDPV_2013} 
seem to be essentially complete now.  In general, one can 
include effects where relevant in the $\gamma$ analyses.  
I note that issues of $CP$ violation and mixing in the $B$ system 
have also been explored, but these are beyond the scope of this review.  

In modes that involve $K_S$, one must also take care with kaon $CP$ 
violation and mixing.  Such effects have been topical recently 
in the context of $\tau$ and $D$ physics 
\cite{BigiSanda_2005, GrossmanNir_2012}; in particular, 
sensitivity to the proper-time acceptance of the $K_S \to \pi^+\pi^-$ decays 
was highlighted.  
More recently, the issues specific to $\gamma$ extractions have 
also been treated \cite{GrossmanSavasito_2014}.  

A further complication arises due to interactions of kaons with material.  
Different $K^0, \bar{K}^0$ interaction rates lead to coherent regeneration, 
as explored in Ref. \cite{Regen_KWGP_2011} in the context $CP$ violation 
searches in $B, D$ decays.   
Further attention is needed here for $\gamma$: 
effects are experiment-specific not only because of acceptance, 
as with the previous issue, but now also due to material details 
as well.  

Another concern involves the required efficiency corrections across the 
$D$ Dalitz plots.  They are used when analyzing charm data 
in order to quote idealized quantities ($R, \delta$, or $c_i, s_i$) with 
efficiency effects removed.  Similarly, different efficiency effects need 
to be accounted for when applying these quantities to the $B$ factory data.  
The issue is the accuracy of such corrections and associated systematic 
uncertainties as overall precision continues to improve.  
This likely deserves further attention.  

Phases are also relevant for $D$ mixing; analyses aim to measure 
the normalized mass and lifetime differences between the physical 
eigenstates: 
$x = \Delta m/\bar{\Gamma}$ and $y = \Delta \Gamma/2\bar{\Gamma}$, 
where $\bar{\Gamma} = (\Gamma_1+\Gamma_2)/2$.  
But the most powerful analyses use hadronic final states with both 
CF and DCSD amplitudes.  In the simplest case, $D \to K\pi$, the 
measured parameters are $x,y$ rotated by the relative strong phase.  
In the case of multi-body decays, there is a similar rotation; 
see, for example, the BaBar result for $D$ mixing extracted via 
$K^-\pi^+\pi^0$ \cite{BaBarKpipi0}.  Here, the rotation is a more 
complicated Dalitz-averaged effect, analogous to effects we have 
discussed for $\gamma$ analyses.  
Currently, threshold data have not yet been analyzed in a way 
that allows one to easily unrotate such multi-body results.  
However, model-independent methods 
are discussed\footnote{Thanks to T. Gershon for pointing out to me, 
at this Workshop in Vienna, that this paper contains such a discussion.} 
in Ref. \cite{BPV_2010}.  

Attention should also be paid to a recent illustration \cite{HR_2014} 
of the potential to extract coherence factors from $D$ mixing analyses 
based on samples taken along with $B$ physics data by both 
$e^+e^-$ and hadron colliders.  Such results can complement, 
or even surpass, charm threshold data.

\Acknowledgements
The author would like to thank his colleagues on CLEO-c and BESIII 
for their hard work on results reported here and also his new colleagues 
on BelleII for their interest in this physics.  Special thanks to 
Onur Albayrak, Tim Gershon, Jim Libby, Alexey Petrov, Guy Wilkinson 
and Jure Zupan for helpful conversations and to Christoph Schwanda 
and his team for a very successful workshop.

\end{document}